\documentclass[compress,final,3p,times,11pt]{elsarticle}
\usepackage{bbm}
\usepackage{amsfonts}
\usepackage{amssymb}
\usepackage{amsmath}
\usepackage{color,epstopdf}
\usepackage{mathrsfs}
\usepackage{graphicx,color,epstopdf}
\usepackage{float}
\usepackage{caption}
\usepackage{subcaption}
\usepackage{float}
\usepackage{subcaption}

\usepackage{parskip}

\usepackage{hyperref}
\journal{arXiv}
\begin{document}
%\textcolor{red}{Here are few questions that I would like we consider for better clarity, readability, and likelihood of acceptance of this manuscript?
%\begin{itemize}
%\item Are objective  and rational clearly defined?
%\item Is our innovation clearly reflected?
%\item Are limitations of the study identified?
%\item Future study clearly indicated
%\item Others?
%\end{itemize}
%}

%\clearpage

%The importance of education and technology in stochastic economic growth.	only facts will be written	
%\begin{frontmatter}
	%\begin{frontmatter}
\title{Stochastic bifurcation in economic growth model driven by L\'evy noise}
\author[mymainaddress]{Almaz Abebe}
\ead{almaz.abebe@howard.edu}
\author[mysecondaryaddress]{Shenglan Yuan\corref{mycorrespondingauthor}}
\ead{shenglanyuan@gbu.edu.cn}
\author[mymainaddress,myfourthaddress]{Daniel Tesfay}
\ead{d.tesfay@ur.ac.rw}
\author[mythirdaddress]{James Brannan}
\ead{jrbrn@clemson.edu}
\cortext[mycorrespondingauthor]{Corresponding author}
\address[mymainaddress]{Department of Mathematics, College of Art and Science, Howard University, Washington DC, 20059, USA}
\address[mysecondaryaddress]{Department of Mathematics,  School of Sciences, Great Bay University, Dongguan 523000, China}
\address[myfourthaddress]{Department of Mathematics, Uppsala University, Box 480, 751 06, Uppsala, Sweden}
\address[mythirdaddress]{Department of Mathematical Sciences, Clemson University, Clemson, South Carolina 29634, USA}

\begin{abstract}
This paper enhances the classical Solow model of economic growth by integrating L\'evy noise, a type of non-Gaussian stochastic perturbation, to capture the inherent uncertainties in economic systems. The extended model examines the impact of these random fluctuations on capital stock and output, revealing the role of jump-diffusion processes in long-term GDP fluctuations. Both continuous and discrete-time frameworks are analyzed to assess the implications for forecasting economic growth and understanding business cycles. The study compares deterministic and stochastic scenarios, providing insight into the stability of equilibrium points and the dynamics of economies subjected to random disturbances. Numerical simulations demonstrate how stochastic noise contributes to economic volatility, leading to abrupt shifts and bifurcations in growth trajectories. This research offers a comprehensive perspective on the influence of external shocks, presenting a more realistic depiction of economic development in uncertain environments.	
\end{abstract}
\maketitle
%\end{frontmatter}
%\linenumbers
\section{Introduction}
Economic development, widely studied as a driver of societal progress, remains inextricably linked to sustained economic growth and enhanced productivity--a relationship first articulated by classical economists in the 18th century. Modern growth theory formalizes these dynamics through mathematical frameworks like the Solow growth model, a seminal contribution by Robert Solow \cite{book} based on his analysis of the U.S. economy. The model explains long-term expansion through the relationship between output and capital accumulation, labor force growth, and technological progress, while assuming diminishing returns to capital. This framework predicts that economies converge toward a steady-state equilibrium where capital and output stabilize in the absence of technological advances, offering critical insights into growth convergence and stability mechanisms \cite{bbb}.

The classical Solow model \cite{S16} provides fundamental insights into economic growth but assumes deterministic and predictable pathways, which clashes with real-world uncertainties such as abrupt policy changes (e.g., tax reforms), technological disruptions, and volatile markets (e.g., financial crises). To address this limitation, researchers have introduced stochastic elements and random variables that simulate unpredictability within the Solow framework. This extension facilitates analysis of how economies adapt to external shocks and improves growth projections under dynamic conditions \cite{asz,A, AG, BS}.

Traditional stochastic models typically rely on Gaussian noise, which captures small, continuous fluctuations but fails to model large, abrupt economic shifts like market crashes or technological breakthroughs. In contrast, L\'evy noise, a non-Gaussian alternative, incorporates jump-diffusion processes. This approach accounts for both gradual variations and sudden extreme events, providing a framework better aligned with real-world volatility. Embedding L\'evy noise into the Solow model enables more effective analysis of economic stability.

This analysis reveals how economies navigate and recover from disruptive shocks in unpredictable environments. The present study examines the contributions of the Solow model to economic growth research, focusing on its application in assessing development parameters such as education, health, and institutional quality. By highlighting the model's core properties within a Cobb-Douglas framework, we analyze  growth dynamics in the absence of exogenous technical or institutional advancements, concentrating on capital and labor as primary drivers.

This study employs both deterministic and stochastic Solow growth models to analyze stochastic bifurcation and economic development, using Gross Domestic Product (GDP) as the primary measure of economic performance. GDP serves as a key indicator of income generation, production, and sectoral output. By treating economic growth as a dependent variable \cite{S11}, the framework highlights its role in enabling governments to enhance public goods--such as education, healthcare, and infrastructure--while underscoring the long-term importance of savings and investment \cite{12}. These mechanisms allocate resources over time through gross capital formation, which includes fixed assets and inventory changes.

Economic development strategies prioritize job creation and improved business environments to foster entrepreneurship and expand societal opportunities \cite{V}. Policies promoting new businesses formation stimulate employment, which is critical for social cohesion; as full employment mitigates public dissatisfaction and aligns economic growth with equitable resource access. Simultaneously, enhanced regulatory frameworks ensure sustainable private-sector expansion, directly linking economic vitality to societal well-being \cite{BSV, FT}.

Productivity, defined as the efficient conversion of inputs (e.g., labor, capital) into outputs (goods and services), underpins economic performance. Technical efficiency--maximizing output from given inputs--quantifies this relationship, with inefficiency reflecting gaps between observed and achievable production levels \cite{xc}. Economists analyze these dynamics through production functions, which map input combinations to maximum possible outputs under existing technology. Two primary methodologies estimate such frontiers: Data Envelopment Analysis (DEA), which is nonparametric and deterministic, and Stochastic Frontier Analysis (SFA), which is parametric and accounts for random shocks. Each offers distinct insights into efficiency rankings and operational benchmarks \cite{R}.

\section{ Solow Economic Growth Model}
\subsection{Deterministic dynamics}
Production transforms inputs (such as labor, capital, and technology) into goods and services. The distribution of output into wages, profits, and taxes then shapes income dynamics and disparities. The Solow model analyzes these input-output transformations, focusing on the steady-state level of capital per worker. In this framework, long-term growth paths are determined by savings, depreciation, and technological progress.

The Cobb-Douglas production function \(Y = AK^\alpha L^{1-\alpha}\)
represents the relationship between output (\(Y\)) and the inputs of capital (\(K\)) and labor (\(L\)). The model includes a measure of technological efficiency \(A\), the output elasticity of capital \(\alpha\), and the output elasticity of labor \(1-\alpha\). This function assumes constant returns to scale.

The rate of change in the capital accumulation equation with the savings rate \(s\) and the capital depreciation rate is modeled by the equation \(dK/dt = s \cdot AK^\alpha - \delta K\). By expressing this  in terms of capital per effective worker (\(k = K / L\)), the fundamental equation of the Solow-Swan  model becomes:
\begin{equation}\label{SS model}
 \frac{d{k}}{dt} = {s} {f}({k}) - (n + g + \delta){k}.
\end{equation}
This model describes the law of motion for capital per effective worker (\(k\)), where
\(sf(k)\) represents savings invested in production, and \((n+g+\delta)k\) accounts for the dilution and depreciation of capital due to population growth (\(n\)), technological growth (\(g\)), and the depreciation rate (\(\delta\)).

 The steady-state level of capital per effective worker (\(k_e\)) in the Solow model (\ref{SS model}) is derived by setting the capital accumulation equation  to zero, i.e., \(\frac{dk}{dt} = 0\):
\begin{equation}\label{steady state capital}
f(k_e) = \frac{n + g + \delta}{s} k_e.
\end{equation}
Using the Cobb-Douglas production function in its intensive form,
\begin{equation}
f(k_e) = \beta k_e^\alpha,
\end{equation}
where \(\alpha\) is the capital share of output. Substituting the above equation into (\ref{steady state capital}), we solve for
\begin{align}
& k_e = \left( \frac{s}{n + g + \delta} \right)^{\frac{1}{1-\alpha}},\;\quad \mbox{assuming}\; k_e > 0 \notag.
\end{align}
The production function can be generalized with a saving rate \(0 < s < 1\) (for instance,
parameterized as \(s = (c_1 + c_2)/2\), a return parameter \(\beta\), and a depreciation rate \(\rho\), yielding
\begin{equation}
k_e = \left( \frac{(c_1 + c_2)\beta}{2\rho} \right)^{\frac{1}{1-\alpha}}.    \notag
\end{equation}
In the steady state, the economy exhibits balanced growth: capital and output grow at a constant rate determined by the rates of savings, depreciation, technological progress and population growth. Deviations from this state lead to transitional dynamics as the economy converges back to \(k_e\). The goal of a steady-state analysis is to identify these equilibrium points, which can be either i)
 a stationary trajectory, where key variables are constant over time,  or ii) a balanced growth path, where all variables grow at the same constant rate.  For more details, see Brannan \cite{bra}.

\subsection{Stochastic dynamics}

The stochastic dynamics in the Solow model of economic growth introduces randomness into the classical framework to account for uncertainties in economic systems, such as technological shocks, policy changes, and other unforeseen events that affect capital accumulation and output. This extension enhances the model's realism by reflecting the inherent unpredictability of real-world economies.

Introducing stochastic noise, such as L\'evy noise, into the Solow model, the system's dynamics become more complex. L\'evy noise captures random, potentially large and abrupt changes, representing shocks like market crashes or technological breakthroughs. This modification incorporates non-Gaussian noise, specifically L\'evy processes, which model random jumps. These jumps can displace the system from its equilibrium path, resulting in periods of economic expansion or contraction. The system now incorporates both continuous fluctuations (represented by Gaussian noise) and discrete jumps (modeled by a L\'evy process), enabling a more comprehensive and realistic depiction of economic growth \cite{as0}. We formulate the model using stochastic differential equation (SDE) with both Gaussian and jump-diffusion components.

The capital stock per unit of effective labor, denoted by \( k(t) \), evolves according to the following SDE:
\begin{equation}\label{Cap}
\frac{dk(t)}{dt} = s(k(t), \gamma) f(k(t)) - \rho k(t) + \sigma dB(t) + \int_{\mathbb{R}} \epsilon(k(t), y) \, N(dt, dy),
\end{equation}
where the interplay of deterministic and stochastic forces shapes its trajectory.
The deterministic component combines a saving rate \( s(k(t), \gamma) \) = $\left[ s_1 + \frac{s_2 - s_1}{1 + e^{-{\gamma} (k - \phi)}} \right]$, where $0 < s_1 < s_2 < 1$  and $\phi$ is a constant reference value of k corresponding to an equilibrium point. This rate \( s(k(t), \gamma) \)
 reflects investment behavior tied to current capital levels and a parameter \( \gamma \).
 Output is modeled by the production function \( f(k(t)) = Bk(t)^{\alpha} \) , with constants
 \( B > 0 \) and \( \alpha \) representing output elasticity.
 Depreciation erodes capital at rate \( \rho \). Stochasticity arises from two sources: continuous Gaussian noise via the Brownian motion term \( \sigma dB(t) \), which captures market volatility, and discontinuous L\'evy jumps represented by \( \int_{\mathbb{R}} \epsilon(k(t), y) N(dt, dy) \). Here, the Poisson random measure \( N(dt, dy) \) and the jump intensity \( \epsilon(k(t), y) \) quantify sudden, unpredictable shifts in capital driven by external shocks or innovations. Together, these elements encapsulate the complex balance of growth, decay, and randomness inherent in economic systems.

The dynamics of investment in economic systems are inherently shaped by time delays and randomness, phenomena rigorously captured by the following SDE:
\begin{equation}\label{Inv}
\frac{dI(t)}{dt} = \beta k(t) I(t) - (v + \gamma) I(t) + \lambda dB(t).
\end{equation}
The growth is propelled by the capital-dependent rate \( \beta k(t) \), counterbalanced by decay parameters \( v \) (depreciation) and \( \gamma \) (attrition), and perturbed by \( \lambda \)-scaled Gaussian noise \( dB(t) \), capturing stochastic market fluctuations. The time delay between investment and its realization as output is implicitly accounted for, as the current investment affects future capital stock and output. The stochastic nature of this system is captured by the noise terms that represent random fluctuations in investment decisions.
% L\'evy Process Representation

To encapsulate both gradual fluctuations and abrupt discontinuities in economic dynamics, we model the system through a L\'evy-driven SDE:
\begin{equation}\label{levpro}
\frac{dX(t)}{dt} = -\eta_a X(t) + \sqrt{\rho \sigma} d(Bt) + \int_{\mathbb{R}} \epsilon(X(t), y) \, N(dt, dy).
\end{equation}
The mean-reverting term \( -\eta_a X(t) \) attenuates past effects at rate \( \eta_a \), the continuous fluctuations are modeled by Gaussian noise \( \sqrt{\rho \sigma} \, dB(t) \), and discontinuous shocks are captured via the L\'evy jump integral \( \int_{\mathbb{R}} \epsilon(X(t), y) N(dt, dy)  \), encoding sudden systemic shifts \cite{lev}.
%Mathematical Analysis of Stochastic Solow Model

We now establish the existence, uniqueness, and stability of  solutions for the system under L\'evy noise.
Equations (\ref{Cap}), (\ref{Inv}), and (\ref{levpro}) augment the classical Solow framework (Eq. \ref{sde1}) by introducing stochastic dynamics, notably L\'evy noise, to account for abrupt, large-scale economic shocks that exceed the modeling capacity of Gaussian noise. This extension bridges deterministic growth theory with real-world volatility, capturing both continuous fluctuations and discontinuous systemic disruptions \cite{YZD,TTYJD,YLZ}, as described by the following system:
\begin{align}\label{sde1}
\frac{dk(t)}{dt} &= s(k(t), \gamma) f(k(t)) - \rho k(t) + \sigma dB(t) + \int_{\mathbb{R}} \epsilon(k(t), y) \, N(dt, dy), \nonumber  \\
\frac{dI(t)}{dt} &= \beta k(t)I(t) - (v + \gamma)I(t) + \lambda dB(t), \\ \nonumber
\frac{dX(t)}{dt} &= -\eta_a X(t) + \sqrt{\rho \sigma} dB(t) + \int_{\mathbb{R}} \epsilon(X(t), y) \, N(dt, dy).
\end{align}
To ensure the system (\ref{sde1}) has a well-defined solution, we verify that it satisfies standard existence and uniqueness criteria for SDEs.
\begin{enumerate}
\item \textbf{Lipschitz Continuity:}
A function \( g(x) \) is Lipschitz continuous if there exists a constant \( L > 0 \) such that:
\begin{equation}
 |g(x_1) - g(x_2)| \leq L|x_1 - x_2|, \quad \forall\, x_1, x_2 \in \mathbb{R}.   \notag
\end{equation}
This condition guarantees that the rate of change of \( g(x) \) remains bounded, preventing  solution trajectories from diverging too rapidly.
\item \textbf{Linear Growth Condition:}
A function \( g(x) \) satisfies the linear growth condition if there exists a constant \( C > 0 \) such that:
\begin{equation}
|g(x)|^2 \leq C(1 + |x|^2), \quad \forall\, x \in \mathbb{R}.    \notag
\end{equation}
\end{enumerate}
This ensures that \( g(x) \) does not grow faster than linearly as \( x \to \infty \), which helps to maintain bounded solutions.

We verify the Lipschitz continuity and linear growth conditions for each term in the SDEs through the following analysis.

1. \textbf{Drift Terms}:  The drift terms in the equations are
\begin{equation}
s(k(t), \gamma)f(k(t)) - \rho k(t), \quad \beta k(t)I(t) - (\nu + \gamma)I(t), \quad -\eta_a X(t).   \notag
\end{equation}
Under the assumption that \( s(k, \gamma) \) and \( f(k) \) are smooth functions, Lipschitz continuity is guaranteed by their bounded derivatives. Furthermore, the linear dependence on \( k(t) \) and \( I(t) \) inherently satisfies the linear growth condition, as their magnitudes scale proportionally with the state variables:
\begin{equation}
|g(k)| \leq C(1 + |k|), \quad |g(I)| \leq C(1 + |I|). \notag
\end{equation}
2. \textbf{Diffusion Terms}:  The diffusion terms in the equations are
\begin{equation}
\sigma dB(t), \quad \lambda dB(t), \quad  \sqrt{\rho \sigma} dB(t). \notag
\end{equation}
The constant coefficients \( \sigma \) and \( \lambda \) encode Gaussian noise. They satisfy Lipschitz continuity and the linear growth condition. Since
\begin{equation}
|c_1-c_2| = 0, \quad \forall\,\, c_1, c_2 \in \mathbb{R}, \notag
\end{equation}
The constant coefficients inherently satisfy linear growth:
\begin{equation}
|g(x)|^2 \leq C(1 + |x|^2), \quad C \text{\; is\; a\; constant}. \notag
\end{equation}
3. \textbf{Jump Terms}:  The jump terms involve a Poisson random measure:
\begin{equation}
\int_{\mathbb{R}}\epsilon(k(t), y)N(dt,dy), \quad  \int_{\mathbb{R}}\epsilon(X(t),y)\, N(dt, dy). \notag
\end{equation}
The jump coefficients \(\epsilon(k, y)\) and \(\epsilon(X(t), y)\) satisfy the Lipschitz condition if
\begin{equation}
 |\epsilon(k_1, y) - \epsilon(k_2, y)|  \leq L|k_1 - k_2|\quad \;\mbox{for\;all}\quad\;  k_1, k_2 \in \mathbb{R},
\end{equation}
ensuring continuity in \(k\) and \(X\). Simultaneously, the linear growth condition holds provided
\begin{equation}
\int_{\mathbb{R}} |\epsilon(k,y)|^2\nu(dy) < C(1+|k|^{2}),
\end{equation}
where \(\nu(dy)\) is the L\'evy intensity measure of \(N(dt,dy)\).

By the existence and uniqueness theorem for SDEs, the system admits a unique strong solution on \([0,T]\) if its coefficients satisfy Lipschitz continuity and linear growth. For the drift terms, these conditions are verified directly through component-wise analysis, while the diffusion terms, being constant, trivially meet both criteria. The jump terms, under the stated assumptions on \(\epsilon(k, y)\), also satisfy the conditions. This well-posedness is rigorously established via foundational stochastic calculus tools, including It\^{o}'s lemma and the Yamada-Watanabe theorem, ensuring robustness against both continuous noise and discontinuous L\'evy-driven shocks.

\subsection{Model stability analysis}

The stability of the economic system depends on the threshold parameter \(\xi\), which determines whether the system converges to a stable growth path or exhibits persistent fluctuations. This threshold is defined as
\begin{equation}
\xi = \frac{\beta B}{\rho (1 + \gamma)}, \notag
\end{equation}
where $\beta$ is the investment rate, $B$ is a constant in the production function, $\rho$ is the depreciation rate, and $\gamma$ represents the intensity of external shocks or perturbations.

When the parameter $\xi>1$, the system tends  to stabilize around the balanced growth path, characterized by regular, bounded economic fluctuations that reflect an inherent resilience to disturbances. Conversely, if $\xi < 1$, persistent negative excursions or instability may emerge, preventing the economy from reverting to this equilibrium path. To further examine the dynamics near this critical threshold, we explore how L\'evy noise, a source of discontinuous and heavy-tailed shocks, influences the capital accumulation process and economic equilibrium. By analyzing the probability distribution of jumps induced by such noise, we highlight the destabilizing role of large, abrupt shocks, which can disrupt growth trajectories even in systems that are nominally stable under smaller disturbances. Notably, at high noise intensities, these extreme jumps amplify the risk of long-term fluctuations and instability, as the economy struggles to absorb recurrent disruptions. This can potentially destabilize the equilibrium and cause sustained deviations from the balanced growth path.
%\subsection{Stability of the Equilibrium Points}

To analyze the stability of the system's equilibrium points, we begin by defining $k^*$ as the equilibrium solution of the capital accumulation process. The stability of $k^*$ is  determined by evaluating the Jacobian matrix of the system at the equilibrium point, which provides critical insights into the local dynamics and resilience of the economic equilibrium. For the system of equations  (\ref{sde1}), the Jacobian matrix is given by
\begin{equation}
J = \begin{pmatrix}
\frac{\partial}{\partial k} \left( s(k, \gamma) f(k) - \rho k \right) & 0 & 0 \\
0 & \frac{\partial}{\partial I} \left( \beta k I - (v + \gamma) I \right) & 0 \\
0 & 0 & \frac{\partial}{\partial X} \left( -\eta_a X \right)
\end{pmatrix}. \notag
\end{equation}
The stability of the equilibrium is determined by the eigenvalues of this Jacobian matrix. If the real parts of all eigenvalues are negative, the equilibrium is stable, meaning that small perturbations will decay over time and the system will return to equilibrium. Conversely, if any eigenvalue has a positive real part, the equilibrium is unstable, and small perturbations will cause the system to deviate further from it.
%\subsection{Lyapunov Exponent Analysis}

To comprehensively assess the system's stability under L\'evy noise, we compute its Lyapunov exponents. These provide a quantitative measure of the exponential divergence (or convergence) between initially adjacent trajectories, where negative values indicate stability (trajectories coalesce) and positive values signal instability (trajectories diverge). For the dynamical system governed by equations (\ref{sde1}), we numerically derive these exponents by simulating the evolution of capital stock and investment processes across varying L\'evy noise intensities. Our analysis reveals a critical trend: as noise intensity escalates, the dominant Lyapunov exponent transitions toward less negative values, reflecting diminished stability and heightened vulnerability to destabilizing large-amplitude fluctuations. This degradation underscores how escalating discontinuous shocks erode the system's capacity to maintain equilibrium, even in regimes that are nominally resistant to smaller perturbations. 	
%\textcolor[rgb]{1.00,0.00,0.00}{ The stochastic version of models  (\ref{1}) may be constructed by adding \emph{Gaussian noise} or \emph{non-Gaussian noise} or \emph{both noises} to $f(X)$. We add noise to the specific rate which emerge representing environmental fluctuations which chiefly affect growth or reproduction of individuals belonging to the population.}
	
\section{Bifurcation Analysis}
Significant understanding of the system's behavior can be gained from the following equations:
\begin{align}\label{kt}
\frac{dk}{dt}&=\beta z^{a}-(\beta-1)p-k, \\ \label{zt}
\varepsilon \frac{dz}{dt}&=-\kappa(z-k), \\ \label{pt}
\varepsilon \frac{dp}{dt}&=-p+\big[1-s(z,\gamma)\big]\frac{\beta}{\beta-1}z^{a}+X(t), \\ \label{Xt}
dX(t)&=-aX(t)dt+\sigma dL^{\alpha}(t),
\end{align}
where $\varepsilon=\eta/\nu$ and $\kappa=\mu/\nu$. When $\varepsilon$ is much smaller than 1, that is, $0<\varepsilon\ll1$, which occurs when the rate constant $\eta=n+g+\delta$ is small compared to $\nu$ and $\mu$ assumed to be of comparable magnitude, the analysis of system \eqref{kt}-\eqref{Xt} simplifies.
For instance, if $1/\eta$ represents the average length of recent business cycles (around 10 years), giving it a nominal value of 0.1, and the average lag time between investment and output is half a year, giving $\nu$ a nominal value of 2, then $\varepsilon$ has a nominal value of 0.05. In such cases, the system of the equations \eqref{kt}-\eqref{Xt} is a slow-fast dynamical system where the capital stock $k$ evolves
more slowly than the variables $z$ and $p$, which adjust rapidly due to the small parameter $\varepsilon$.

\begin{figure}
\centering
\includegraphics[width=10cm]{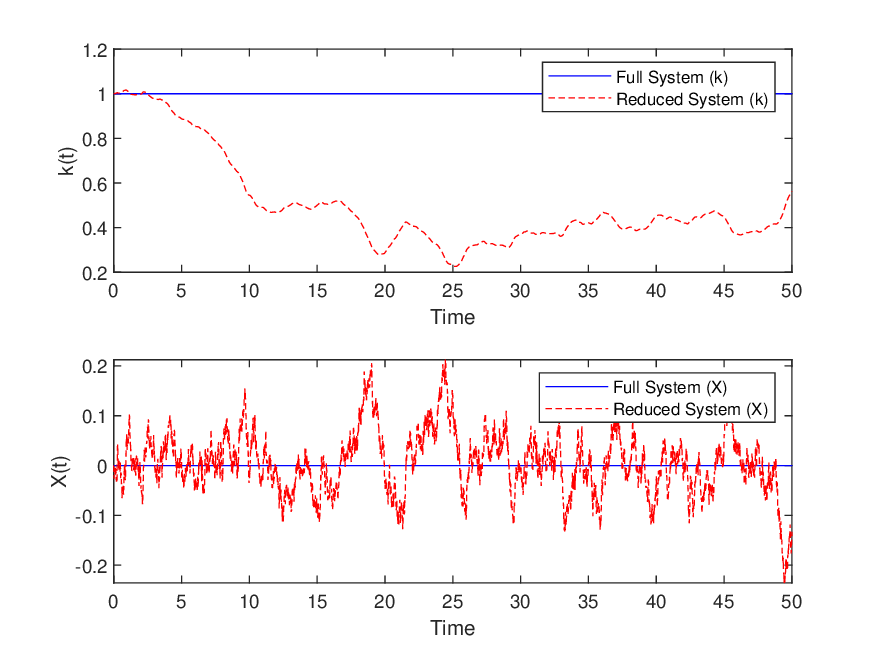}
\caption{The time series of both the system with $\varepsilon=0.05$ and the reduced system.}\label{k}	
\end{figure}

Approximating the solutions to the equations \eqref{zt}-\eqref{pt} for $z$ and $p$ by
\begin{equation}\label{varepsilon0}
z=k+O(\varepsilon)\quad\text{and}\quad p=[1-s(k,\gamma)]\frac{\beta}{\beta-1}k^{a}+X(t)+O(\varepsilon)\quad\text{as}\,\,\, \varepsilon\rightarrow0,
\end{equation}
and ignoring $O(\varepsilon)$ terms, we substitute these approximations for $z$ and $p$ as in Eqs. \eqref{varepsilon0} into the equation \eqref{kt} for $k$. This yields the slow time approximation to system \eqref{kt}-\eqref{Xt}:
\begin{align}\label{slow}
\frac{dk}{dt}&=\big[\beta s(k,\gamma)k^{a}-k\big]-(\beta-1)X(t), \\ \label{noise}
dX(t)&=-aX(t)dt+\sigma dL^{\alpha}(t).
\end{align}
Thus, by approximating $z\approx k$ and $p\approx[1-s(k,\gamma)]\frac{\beta}{\beta-1}k^{a}+X(t)$, the system \eqref{kt}-\eqref{Xt} reduces to a slow equation \eqref{slow} for $k$ coupled with an Ornstein-Uhlenbeck process $X(t)$ characterized by \eqref{noise}. Assuming parameter values $s_1=0.2$, $s_2=0.8$, $\beta=2$, $a=0.3$, $\kappa=1$, $\sigma=0.1$, $\alpha=1$. Fig. \ref{k} compares the full system with $\varepsilon=0.05$ and the reduced slow system.

\begin{figure}
\centering
\includegraphics[width=10cm]{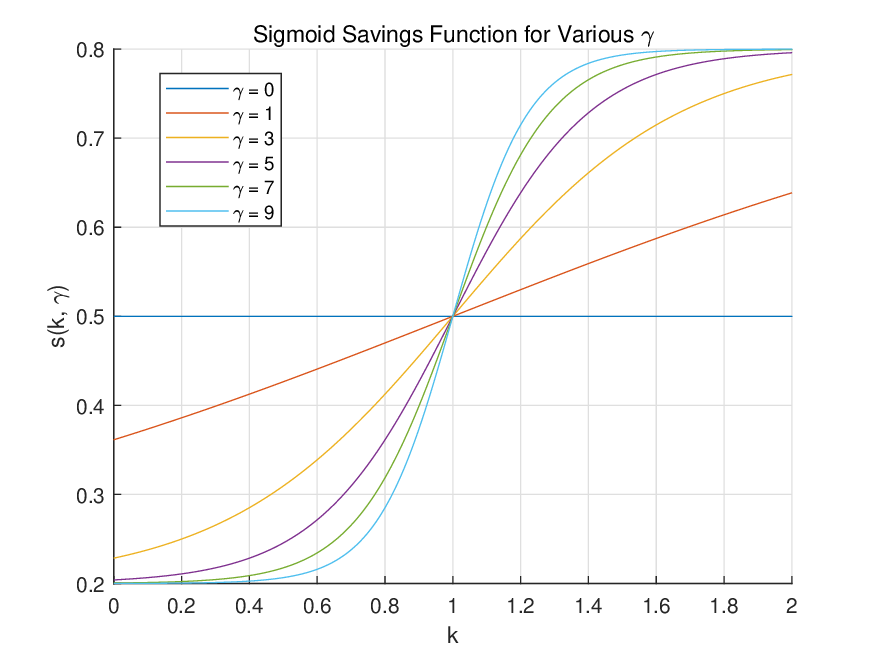}
\caption{The sigmoid savings function $s(k,\gamma)$.}\label{SSF}	
\end{figure}

Note that in slow time, the equality $z=k$ indicates that the effect of the time delay is negligible.
When $\gamma=0$, the slow equation \eqref{slow} reduces to the dimensionless Solow model with a constant saving rate
$\beta s(k,0)=1$,  disturbed by an Ornstein-Uhlenbeck process:
\begin{equation*}
\frac{dk}{dt}=k^{a}-k-(\beta-1)X(t).
\end{equation*}
To analyze how $s(k,\gamma)$ affects the behavior of solutions of Eq. \eqref{slow} in the absence of noise, we consider the deterministic case ($X(t)=0$). The dynamics of $k$ are then governed by
\begin{equation}\label{kwn}
\frac{dk}{dt}=\beta s(k,\gamma)k^{a}-k.
\end{equation}
Here, $s(k,\gamma)=s_1+\frac{s_2-s_1}{1+e^{-\gamma(k-1)}}$ is a sigmoid function transitioning between savings rates $s_1$ and $s_2$, where $0<s_1<s_2<1$.
As $\gamma$ increases from zero, the graphs of $s(k,\gamma)$ for several values of $\gamma$ are shown in Fig. \ref{SSF}.

\begin{figure}
\centering
\includegraphics[width=10cm]{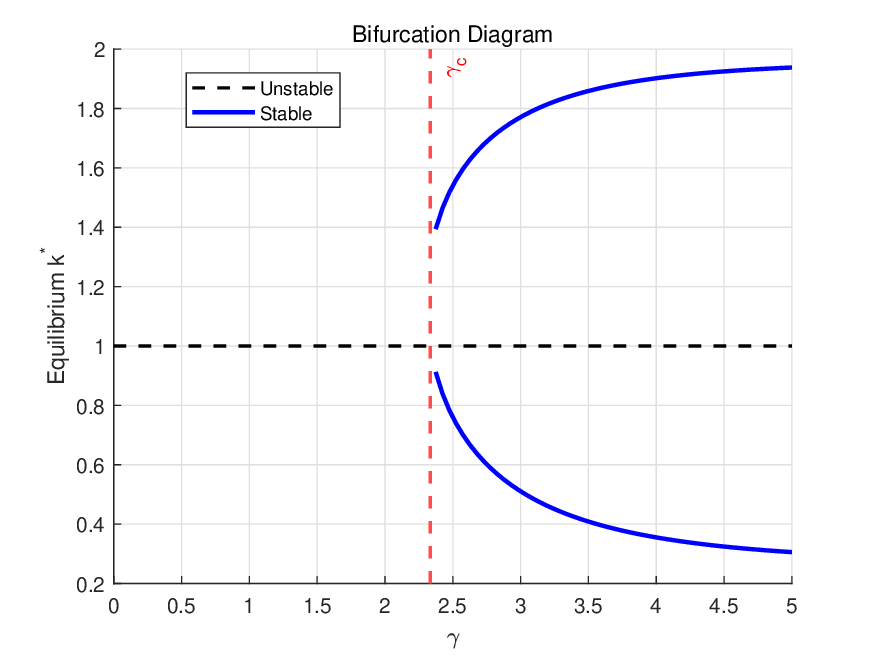}
\caption{The bifurcation diagram shows equilibrium solutions $k^{\ast}$ vs $\gamma$.
The equilibrium $k_0^{\ast}=1$ exists for all $\gamma\geq0$, provided $s(1,\gamma)=1/\beta$.
A critical value $\gamma_c\approx2.3333$ marks a bifurcation. For $\gamma<\gamma_c$, $k_0^{\ast}=1$ is stable. For
$\gamma>\gamma_c$, two new stable equilibria $k_{+}^{\ast}$ and $k_{-}^{\ast}$emerge, and $k_0^{\ast}$ becomes unstable.
}\label{BifurcationDiagram}	
\end{figure}

All equilibrium solutions of Eq. \eqref{kwn} are found by setting the right-hand side of Eq. \eqref{kwn} to zero, resulting in the nonlinear equation
\begin{equation}\label{0kwn}
\beta s(k,\gamma)k^{a}-k=0.
\end{equation}
One solution of Eq. \eqref{0kwn} is $k_{0}^{\ast}(\gamma)=1$ for all $\gamma\geq0$, which corresponds to the balanced growth path
in Eq. \eqref{kwn}. For $\gamma$ below a critical value $\gamma_c$, $k_{0}^{\ast}(\gamma)=1$ is the only solution of Eq. \eqref{0kwn}. As $\gamma$ increases past $\gamma_c$, two additional solutions $k_{+}^{\ast}(\gamma)$ and $k_{-}^{\ast}(\gamma)$ of Eq. \eqref{0kwn} appear. Therefore, $(\gamma_c,1)$ is a bifurcation point for Eq. \eqref{kwn}. We solve the equilibrium equation for different $\gamma$ and plot the branches. The bifurcation diagram showing equilibria $k^{\ast}$ vs. $\gamma$ for Eq. \eqref{kwn} is shown in Fig \ref{BifurcationDiagram}. We compute the critical bifurcation parameter $\gamma_c\approx2.3333$. This bifurcation highlights the emergence of multiple stable growth paths, with implications for economic resilience and path dependence.

The bifurcation point \cite{YW} is computed by linearizing the equation \eqref{kwn} near
the equilibrium solution $k_{0}^{\ast}(\gamma)=1$. Letting $k=1+v$, we find that
\begin{equation}\label{vt}
\frac{dv}{dt}=-\Big(1-a-\frac{1}{2}\frac{\psi-1}{\psi+1}\gamma\Big)v+O(u^{2})\quad\text{as}\, \,\,u\rightarrow0,\quad \psi=\frac{s_2}{s_1}.
\end{equation}
Omitting the nonlinear term in Eq. \eqref{vt} gives the linear approximation to Eq. \eqref{kwn} in the neighborhood of the
equilibrium solution $k_{0}^{\ast}(\gamma)=1$:
\begin{equation}\label{v}
\frac{dv}{dt}=-r(\gamma)v,
\end{equation}
where
\begin{equation}\label{r}
r(\gamma)=\Big(1-a-\frac{1}{2}\frac{\psi-1}{\psi+1}\gamma\Big).
\end{equation}
Linearizing around $k_{0}^{\ast}(\gamma)=1$ reveals the stability coefficient $r(\gamma)$. The linearized equation \eqref{v} has solutions of the form
\begin{equation*}
v=v_0e^{-r(\gamma)t}.
\end{equation*}
Assuming $v_{0}\neq0$, if the decay rate constant $r(\gamma)>0$, $v(t)$ decays to zero. On the other hand, when $r(\gamma)<0$, in
which case $-r(\gamma)>0$ is a growth rate constant, $|v(t)|$ increases as $t$ increases. From this, we conclude that the bifurcation
point occurs at the value of $\gamma$ such that $r(\gamma)=0$, yielding the critical bifurcation parameter:
\begin{equation*}
\gamma_{c}=2\frac{\psi+1}{\psi-1}(1-a),
\end{equation*}
which marks the transition to bistability in the economic growth model. We may also write
\begin{equation*}
r(\gamma)=\frac{1}{2}\frac{\psi-1}{\psi+1}(\gamma_{c}-\gamma).
\end{equation*}
Thus, the equilibrium state $k_{0}^{\ast}(\gamma)=1$ is asymptotically stable for $r(\gamma)>0$ if $\gamma<\gamma_{c}$. When $\gamma>\gamma_{c}$, we have $r(\gamma)<0$, indicating instability at $k_{0}^{\ast}(\gamma)=1$.

\begin{figure}
\begin{center}
\begin{minipage}{3.2in}
\leftline{(a)}
\includegraphics[width=3.2in]{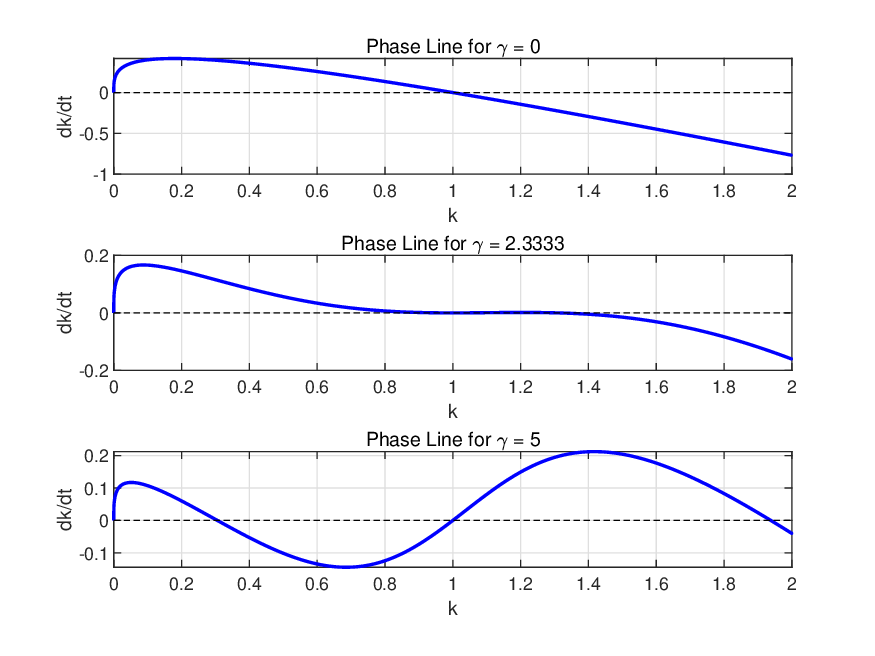}
\end{minipage}
\hfill
  \begin{minipage}{3.2in}
\leftline{(b)}
\includegraphics[width=3.2in]{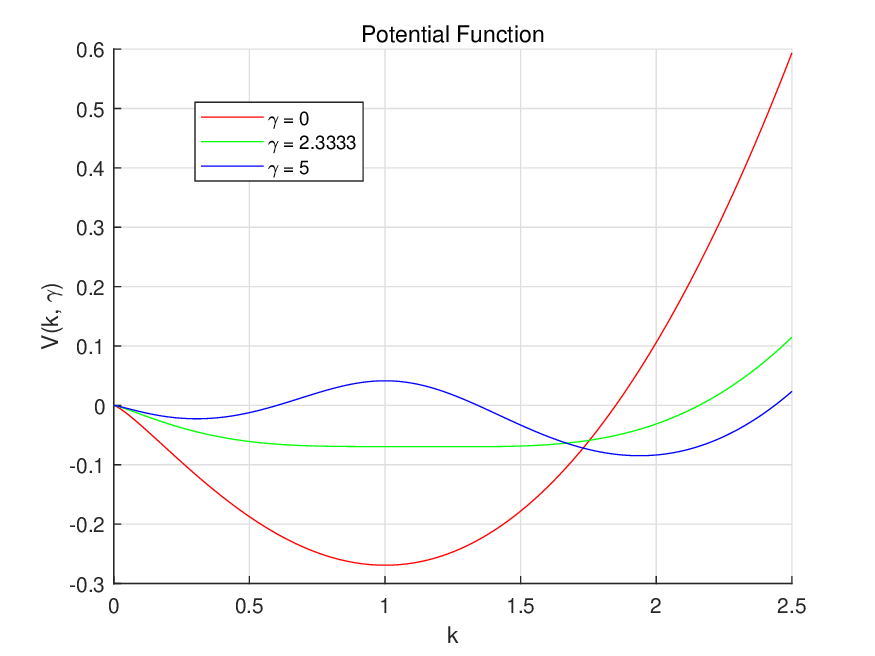}
\end{minipage}
\caption{(a) The phase line plots plot $\frac{dk}{dt}$ against $k$ for three different $\gamma$ values. The system \eqref{kwn} transitions from monostable to bistable as $\gamma$ exceeds $\gamma_c$, driven by the nonlinear savings function $s(k,\gamma)$; (b) The potential function $V(k,\gamma)$ for different $\gamma$, highlighting the double-well when $\gamma>\gamma_c$, illustrates bistability with minima at $k_{+}^{\ast}(\gamma)$ and $k_{-}^{\ast}(\gamma)$, and a maximum at $k_{0}^{\ast}(\gamma)=1$.}\label{PLP2PF1}
\end{center}
\end{figure}

Fig. \ref{PLP2PF1}(a) involves phase line plots of $\frac{dk}{dt}$ against $k$ for different $\gamma$ values, illustrating stable and unstable points. We show phase line plots of Eq. \eqref{kwn} for various values of $\gamma$. These plots confirm that the two equilibria $k_{+}^{\ast}(\gamma)$ and $k_{-}^{\ast}(\gamma)$ that appear for $\gamma>\gamma_{c}$ are stable. Finally, Fig. \ref{PLP2PF1}(b) shows graphs of the potential function
\begin{equation*}
V(k,\gamma)=-\int\big(\beta s(k,\gamma)k^{a}-k\big)dk
\end{equation*}
for various values of $\gamma$. In terms of this potential, the dynamics of Eq. \eqref{kwn} can be written as
\begin{equation*}
\frac{dk}{dt}=-\frac{\partial V}{\partial k}.
\end{equation*}
The potential functions $V(k,\gamma)$ illustrate the stability and behavior of the equilibria for various values of $\gamma$.
If $\gamma>\gamma_{c}$, Fig. \ref{PLP2PF1}(b) shows that $V(k,\gamma)$ has local minima corresponding to the stable equilibria $k_{+}^{\ast}(\gamma)$ and $k_{-}^{\ast}(\gamma)$, and a local
maximum at the unstable equilibrium $k_{0}^{\ast}(\gamma)=1$. The potential in this case exhibits a double-well shape, indicating that the system is bistable, with two stable equilibria separated by an unstable one.

\section{Comparison of Stochastic Models}
To better understand the impact of L\'evy noise on economic growth, we perform numerical simulations of the system described by the stochastic differential equations (\ref{sde1}). These simulations visualize how L\'evy noise influences capital accumulation and output over time. For the purpose of the simulations, we discretize the system using the Euler-Maruyama method, which is a widely used method for solving SDEs numerically.

The Euler-Maruyama method approximates the continuous system by discretizing time into small steps, allowing us to calculate the evolution of variables at each step. For a given time step $\Delta t$, the Euler-Maruyama discretization of the system (\ref{sde1}) is given by:
\begin{equation}
k_{n+1} = k_n + \Delta t \left[ s(k_n, \gamma) f(k_n) - \rho k_n \right] + \sqrt{\Delta t} \sigma \, Z_n + \Delta t \int_{\mathbb{R}} \epsilon(k_n, y) N(dt, dy),
\end{equation}
where $Z_n$ is a standard normal random variable, and the discretized L\'evy jump term is approximated from the integral $\int_{\mathbb{R}}\epsilon(k_n,y)N(dt,dy)$ over the jump distribution $N(dt, dy)$.

We simulate the system over a time period $T$ with a total of $N$ time steps, where each time step corresponds to one quarter (three months). The parameters for the simulation are chosen based on realistic estimates of economic growth parameters:
\begin{equation*}
\alpha=0.33, \quad \beta=0.4, \quad \rho=0.02, \quad \sigma=0.1, \quad \gamma=0.5.
\end{equation*}
For the L\'evy noise, we set the jump intensity and distribution parameters to:
\begin{equation*}
\lambda = 0.05, \quad \text{and} \quad N(dt, dy) \sim \text{Poisson}(0.01).
\end{equation*}
These parameters reflect a scenario where the economy subject to small but significant jumps due to external shocks.

\begin{figure}
    \centering
\begin{subfigure}[b]{0.75\textwidth}
        \centering
        \includegraphics[width=\textwidth]{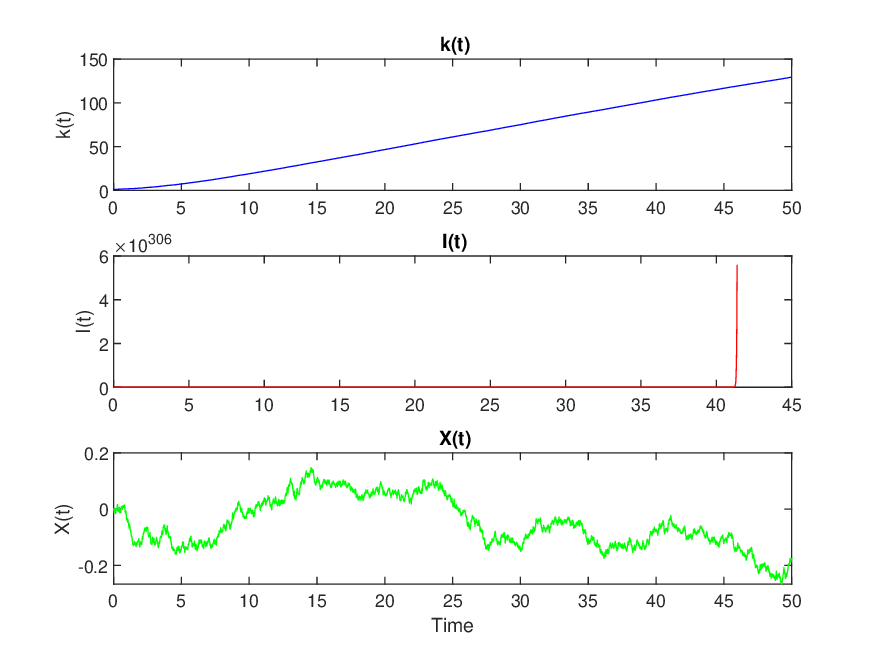}
    \caption{ SDE with Gaussian and L\'evy Noise with $\alpha = 0.33$;}
        \label{SDElevy1}
    \end{subfigure}
    \vspace{0.5em}
    \begin{subfigure}[b]{0.75\textwidth}
        \centering
\includegraphics[width=\textwidth]{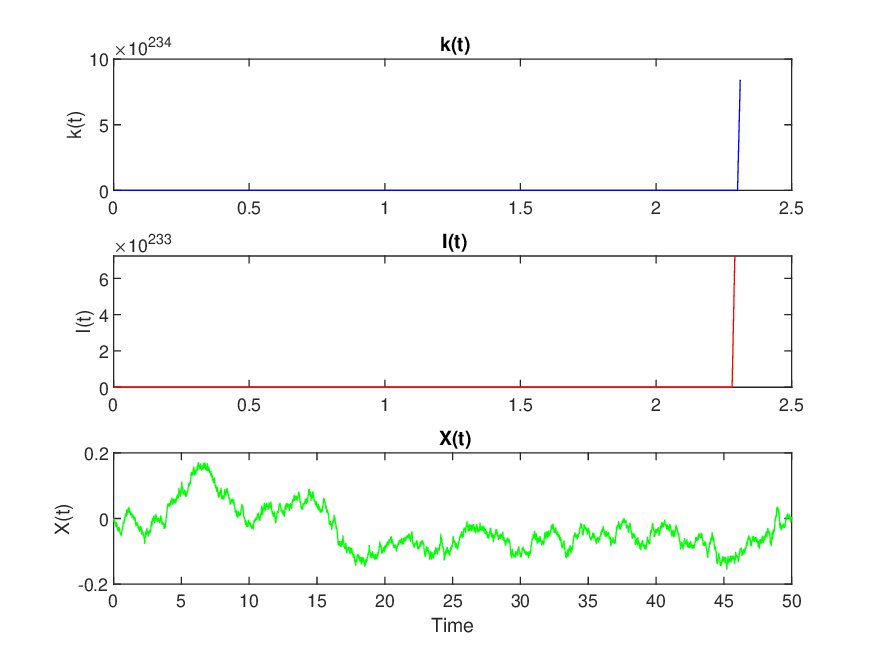}
        \caption{SDE with Gaussian and L\'evy Noise with $\alpha = 1.5$;}
        \label{SDElevy2}
    \end{subfigure}
\caption{Simulations of the stochastic dynamic system for capital per effective worker \( k(t) \), investment \( I(t) \), and the external shock process \( X(t) \). The savings function is sigmoidal, \( s(k) = s_1 + (s_2 - s_1)/(1 + e^{-\gamma(k - \phi)}) \), with \( s_1 = 0.2 \), \( s_2 = 0.8 \), \( \gamma = 0.5 \), and \( \phi = 1.0 \). The production function is \( f(k) = B k^\alpha \) with \( B = 1.5 \). Stochastic components include Brownian motion and compound Poisson jumps with intensity \( \lambda = 0.01 \). Other parameters are: capital depreciation rate \( \rho = 0.02 \), investment decay rate \( v = 0.02 \), capital-investment interaction rate \( \beta = 0.4 \), Brownian noise scale \( \sigma = 0.1 \), and mean-reversion rate \( \eta_a = 0.1 \). The simulation runs over \( T = 50 \) time units with a step size of \( \Delta t = 0.01 \). }
\label{SDEsimulations}
     \vspace{0.5em}
\end{figure}

Fig. \ref{SDEsimulations}  displays a representative simulation of the stochastic dynamic system for capital per effective worker
 \( k(t) \),  investment  \( I(t) \), the external shock process \( X(t) \), governed by stochastic differential equations. The results demonstrate how the interplay of Gaussian diffusion and L\'evy jumps generates complex growth trajectories, characterized by periods of stable growth punctuated by abrupt shifts due to large shocks.

The simulations are run for a period of 50 years with quarterly time steps. Comparing the capital accumulation process under L\'evy noise to the deterministic version of the model without L\'evy noise. The results reveal that the inclusion of L\'evy noise leads to significantly larger fluctuations in capital and output, particularly during periods of high volatility.

The results clearly demonstrate that the inclusion of L\'evy noise introduces irregular jumps in capital stock, resulting in more erratic output fluctuations compared to the deterministic model. These jumps represent significant economic shocks that cannot be captured by Gaussian noise alone.
	
We further compare the results from the L\'evy noise model with one driven by Gaussian noise (Brownian motion) instead of L\'evy noise. This comparison shows that although both models follow similar long-term trends, the L\'evy noise model produces more pronounced deviations from the expected growth path due to the inclusion of large, sudden jumps.

The figure illustrates that the Gaussian noise model results in smooth fluctuations, whereas the L\'evy noise model  leads to abrupt, significant shifts in the capital accumulation process.  This contrast underscores the importance of incorporating jump-diffusion processes in economic modeling, particularly for analyzing volatile environments.

\section{Conclusions}

This paper has enhanced the classical Solow growth model by incorporating L\'evy noise to better capture large, unexpected economic fluctuations. The integration of jump-diffusion processes offers a more accurate representation of economic dynamics, particularly in volatile environments. Future work will include applying this model to empirical data and exploring its implications for economic policy design.

Mathematically, the extended Solow model with L\'evy noise demonstrates that the system can exhibit complex dynamics, including abrupt, large-scale deviations in capital and output. The threshold parameter $\xi$ is critical, determining whether the system converges to a stable growth path or experiences persistent fluctuations. Our analysis further confirms that L\'evy noise can substantially impact system stability, amplifying volatility in economic growth.

Numerical simulations underscore the substantial impact of L\'evy noise on capital accumulation and output. By including jump-diffusion processes, the model more accurately captures the randomness and irregularities inherent in real-world economic data. These findings highlight that incorporating non-Gaussian noise is essential for economic modeling, especially for improving growth forecasts under uncertainty.

The L\'evy-augmented Solow model provides a more nuanced perspective on economic growth by embedding realistic stochastic disturbances into a established framework. The results show that random shocks can significantly alter economic trajectories, leading to pronounced variations in capital stock and aggregate output. A key advantage of L\'evy noise is its capacity to model large, discontinuous changes, thereby revealing the potential for sudden, disruptive shifts in economic stability.

These insights underscore the importance of designing economic policies that account for stochastic factors, thereby enhancing resilience against unexpected shocks. Ultimately, this research affirms the vital role of sophisticated stochastic modeling in achieving a realistic and comprehensive understanding of economic growth in an unpredictable global context.
	
\bigskip

\noindent\textbf{Data Availability}

The numerical algorithms and source code that support the findings of this study are available from the corresponding author upon reasonable request.

\bigskip
\noindent\textbf{Declaration of competing interest}

No author associated with this paper has disclosed any potential or pertinent conflicts which
may be perceived to have impending conflict with this work.

\bigskip
\noindent\textbf{Acknowledgements}

This work was supported by the Guangdong Basic and Applied Basic Research Foundation (Grant No. 2025A1515012560) and the Guangdong Introduction Program (Grant No. 2023QN10X753)

%\renewcommand{\bibname}{References} % change title name bibiliography to references
	%\bibliographystyle{ieeetr}
	%\fontsize{10}{10}\selectfont{\bibliography{references.bib}}
\end{document}